\documentclass[conference]{IEEEtran}
\IEEEoverridecommandlockouts

\usepackage{cite}
\usepackage{amsmath,amssymb,amsfonts}
\usepackage{graphicx}
\usepackage{textcomp}
\usepackage[dvipsnames]{xcolor}
\usepackage[acronym,shortcuts]{glossaries}
\usepackage{bm}
\usepackage{caption}
\usepackage{subfigure}
\usepackage{algorithm, algpseudocode}
\usepackage{algcompatible}
\usepackage{mathtools}
\usepackage{tabularx,flushend}
\def\BibTeX{{\rm B\kern-.05em{\sc i\kern-.025em b}\kern-.08em
T\kern-.1667em\lower.7ex\hbox{E}\kern-.125emX}}

\newcommand{\trans}[0]{^{\mathsf{T}}}
\newcommand{\herm}[0]{^{\mathsf{H}}}

\newcommand{\Real}[1]{\Re\{{#1}\}}

\newacronym{OTFS}{OTFS}{orthogonal time frequency space}
\newacronym{AFDM}{AFDM}{affine frequency division multiplexing}
\newacronym{MIMO}{MIMO}{multiple-input multiple-output}
\newacronym{ISAC}{ISAC}{integrated sensing and communications}
\newacronym{3D}{3D}{three-dimensional}
\newacronym{2D}{2D}{two-dimensional}
\newacronym{RX}{RX}{receiver}
\newacronym{TX}{TX}{transmitter}
\newacronym{mmWave}{mmWave}{millimeter-wave}
\newacronym{ULA}{ULA}{uniform linear array}
\newacronym{QAM}{QAM}{quadrature amplitude modulation}
\newacronym{AWGN}{AWGN}{additive white Gaussian noise}
\newacronym{OFDM}{OFDM}{orthogonal frequency division multiplexing}
\newacronym{BS}{BS}{base station}
\newacronym{DFT}{DFT}{discrete Fourier transform}
\newacronym{TD}{TD}{time-domain}
\newacronym{CP}{CP}{cyclic prefix}
\newacronym{DAFT}{DAFT}{discrete affine Fourier transform}
\newacronym{IDAFT}{IDAFT}{inverse discrete affine Fourier transform}
\newacronym{CPP}{CPP}{\textit{chirp-periodic} prefix}
\newacronym{IDZT}{IDZT}{inverse discrete Zak transform}
\newacronym{DZT}{DZT}{discrete Zak transform}
\newacronym{ICI}{ICI}{inter-carrier interference}
\newacronym{BER}{BER}{bit error rate}
\newacronym{DoF}{DoF}{degrees-of-freedom}
\newacronym{AoD}{AoD}{angle-of-departure}
\newacronym{AoA}{AoA}{angle-of-arrival}
\newacronym{SIM}{SIM}{stacked intelligent metasurfaces}
\newacronym{FIM}{FIM}{flexible intelligent metasurface}
\newacronym{UPA}{UPA}{uniform planar array}
\newacronym{I/O}{I/O}{input-output}
\newacronym{iid}{i.i.d.}{independent and identically distributed}
\newacronym{V2X}{V2X}{vehicle-to-everything}
\newacronym{LEO}{LEO}{low-earth orbit}
\newacronym{RIS}{RIS}{reconfigurable intelligent surface}
\newacronym{DD}{DD}{doubly-dispersive}
\newacronym{LoS}{LoS}{line-of-sight}
\newacronym{6G}{6G}{sixth generation}
\newacronym{FPDD}{FPDD}{FIM-parameterized DD}
\newacronym{MUSIC}{MUSIC}{Multiple Signal Classification}
\newacronym{SNR}{SNR}{signal-to-noise ratio}
\newacronym{QoS}{QoS}{quality of service}
\newacronym{B5G}{B5G}{beyond fifth generation}
\newacronym{UAV}{UAV}{unmanned aerial vehicle}
\newacronym{LMMSE}{LMMSE}{linear minimum mean square error}
\newacronym{ISFFT}{ISFFT}{inverse symplectic finite Fourier transform}
\newacronym{SFFT}{SFFT}{symplectic finite Fourier transform}
\newacronym{IDFT}{IDFT}{inverse discrete Fourier transform}
\newacronym{wlg}{wlg}{without loss of generality}
\newacronym{QPSK}{QPSK}{quadrature phase-shift keying}
\newacronym{SVD}{SVD}{singular value decomposition}
\newacronym{RF}{RF}{radio frequency}
\newacronym{CRB}{CRB}{Cram\'{e}r-Rao bound}

\begin{document}

\title{Bistatic Integrated Sensing and Communications \\ with Flexible Intelligent Metasurfaces}

\author{
\IEEEauthorblockN{Iv\'{a}n Alexander Morales Sandoval$^*$, Thushar Venkataramanaiah$^*$, Kuranage Roche Rayan Ranasinghe$^*$,\\ Jiancheng An$^\dag$, Hyeon Seok Rou$^*$, Giuseppe Thadeu Freitas de Abreu$^*$}
\IEEEauthorblockA{\textit{$^*$School of Computer Science and Engineering, Constructor University, Bremen, Germany} \\
\textit{$^\dag$School of Electrical and Electronics Engineering, Nanyang Technological University, Singapore} \\
Emails: \{imorales, tvenkataramanaiah, kranasinghe, hrou, gabreu\}@constructor.university,
jiancheng.an@ntu.edu.sg}

\thanks{An extended version of this work has been published in the Transactions on Wireless Communications \cite{RanasingheTWCFIM}.}
}

\maketitle

\begin{abstract}
We propose a novel \ac{DD} \ac{MIMO} channel model incorporating \acp{FIM}, suitable for \ac{ISAC} in high-mobility scenarios.
We show how the proposed \ac{FPDD} channel model extends to multicarrier waveforms known to perform well in \ac{DD} environments, namely, \ac{OFDM}, \ac{OTFS}, and \ac{AFDM}.
Leveraging this model, we formulate an achievable rate maximization problem with a sensing constraint for all waveforms and solve it via gradient ascent with closed-form gradients.
Numerical results indicate that \ac{FIM} technology significantly impacts the achievable rate, with careful parametrization essential for strong \ac{ISAC} performance across all waveforms.
\end{abstract}


\glsresetall

\section{Introduction}

Next-generation wireless networks in the \ac{B5G}/\ac{6G} era will increasingly operate in challenging high-mobility multipath environments such as \ac{V2X}, \ac{UAV}, and \ac{LEO} satellite links \cite{GiordaniCOMMAG2020}.
Such environments exhibit \ac{DD} channels that incorporate both Doppler shifts and multipath delays \cite{Bliss_Govindasamy_2013}.
The inherent delay-Doppler structure of these channels can be leveraged for \ac{ISAC} \cite{liu2022integrated}, enabling radar parameter estimation directly from communication waveforms \cite{RanasingheARXIV2024}, and motivating high-mobility waveforms such as \ac{OTFS} and \ac{AFDM} \cite{Rou_SPM_2024}.

A complementary approach is to manipulate the propagation environment using reconfigurable electromagnetic structures.
While \acp{RIS} offer electronically controlled phase shifts \cite{AlexandropolousVTM2024} and \acp{SIM} enable wave-domain processing via stacked meta-atom layers \cite{AnWC2024}, the emerging \ac{FIM} paradigm can physically reconfigure its \ac{3D} surface shape in response to channel conditions \cite{TWC_2024_An_Flexible, Nature_2022_Bai_A}.
Unlike \acp{RIS}/\acp{SIM}, \acp{FIM} adjust meta-atom positions along the normal direction, providing enhanced spatiotemporal control \cite{TCOM_2025_An_Flexible}.
Recent works demonstrated that \acp{FIM} can double \ac{MIMO} capacity \cite{TCOM_2025_An_Flexible} and improve sensing power by $3$\,dB \cite{TVT_2025_Teng_Flexible} compared to rigid arrays.

However, integrating \acp{FIM} with \ac{DD} channel models for high-mobility \ac{ISAC} remains open.
Prior work on metasurface-parameterized \ac{DD} channels \cite{ranasinghe2025metasurfacesintegrateddoublydispersivemimochannel, ranasinghe2025doublydispersivemimochannelsstacked} considered \acp{SIM} and \acp{RIS}, where the element response enters linearly via complex reflection coefficients.
In contrast, \ac{FIM} geometry alters the propagation phase, \acp{AoA}, and \acp{AoD} \emph{nonlinearly} through the array manifold $\mathbf{b}(\bm{y},\phi,\theta)$, introducing geometry coordinates $\bm{y}$ as a new optimization variable absent in prior works.

In light of this background, we introduce a novel \ac{FPDD} channel model in which the spatial array response depends explicitly on the \ac{FIM} geometry, naturally extending to \ac{OFDM}, \ac{OTFS}, and \ac{AFDM} waveforms with complete \ac{I/O} relationships.
Then, we formulate and solve an achievable-rate maximization problem under a sensing constraint using gradient ascent with closed-form shape-gradient expressions that capture the sensitivity of mutual information to \ac{FIM} geometry. These gradients vanish for non-morphing metasurfaces, highlighting the new \ac{DoF} from \acp{FIM}.
Finally, simulations demonstrate that geometry morphing substantially impacts achievable rate and sensing quality, producing gains unattainable with non-morphing metasurfaces.

\textit{Notation:}
Column vectors and matrices are denoted by bold lowercase and uppercase letters, respectively.
$\mathbf{A}\trans$, $\mathbf{A}\herm$, and $[\mathbf{A}]_{i,j}$ denote the transpose, Hermitian, and $(i,j)$-th element of $\mathbf{A}$.
$\mathbf{I}_N$ and $\mathbf{F}_N$ are the $N\times N$ identity and normalized \ac{DFT} matrices, $\otimes$ denotes the Kronecker product, $\odot$ the Hadamard product, and $\jmath\triangleq\sqrt{-1}$.

\section{The Proposed FIM MIMO Channel Model}
\label{FIM_MIMO_Model}

We consider a point-to-point \ac{MIMO} system employing \acp{FIM} at both the transmitter and receiver, with $P$ scatterers in the environment.
The \ac{TX}-\ac{FIM} transmits a waveform (\ac{OFDM}, \ac{OTFS}, or \ac{AFDM}) used for both communication and probing of scatterers; the \ac{RX}-\ac{FIM} performs joint data reception and \ac{AoA} estimation.

\subsection{FIM Antenna Array Response}

Consider a flexible \ac{UPA} with $B \triangleq B_x B_z$ elements, where $B_x$ and $B_z$ are the number of elements along the $x$- and $z$-axes with spacing $d_x = d_z = \lambda/2$ (half-wavelength to avoid spatial aliasing~\cite{AnTWC2025}), and $\lambda$ is the carrier wavelength.
Let $\bm{p}_b = [x_b, y_b, z_b]\trans \in \mathbb{R}^3$ be the location of the $b$-th element for $b \in \mathcal{B} \triangleq \{1,\dots,B\}$, with
\begin{subequations}
\begin{equation}
x_b = d_x \!\times\! \text{mod}(b\!-\!1,B_x),\quad
z_b = d_z \!\times\! \lfloor (b\!-\!1)/B_x \rfloor.
\end{equation}
\end{subequations}

The key principle of \acp{FIM} is that each element can move along the $y$-axis (normal to the planar aperture) within 
\begin{equation}
y_\text{min} \leq y_b \leq y_\text{max},\quad \forall b \in \mathcal{B},
\label{eq:morph_constraint}
\end{equation}
with morphing range $\zeta \triangleq y_\text{max} - y_\text{min} > 0$.

The $x$- and $z$-coordinates remain fixed, requiring no additional mechanical complexity.
The \ac{FIM} shape is then fully characterized by
\begin{equation}
    \bm{y} \triangleq [y_1, y_2, \dots, y_B]\trans \in \mathbb{R}_+^{B \times 1}.
\end{equation}

For elevation angle $\theta\in[0,\pi]$ and azimuth angle $\phi\in[-\frac{\pi}{2},\frac{\pi}{2}]$, the \ac{UPA} response vector $\mathbf{b}(\bm{y},\phi,\theta) \in \mathbb{C}^{B \times 1}$ is
\begin{align}
\!\!\mathbf{b}(\bm{y},\phi,\theta) &\triangleq \tfrac{1}{\sqrt{B}} \Big[ 1, e^{\jmath \frac{2\pi}{\lambda} ( x_1 \sin\theta \cos\phi + y_1 \sin\theta \sin\phi + z_1 \cos\theta )},  \nonumber \\
&\hspace{-2ex} \dots, e^{\jmath \frac{2\pi}{\lambda} ( x_B \sin\theta \cos\phi + y_B \sin\theta \sin\phi + z_B \cos\theta )} \Big]\trans.
\label{eq:steering_vec_FIM}
\end{align}

Critically, the $y_b$ coordinates enter the phase terms \emph{nonlinearly} through the product $y_b \sin\theta \sin\phi$, coupling the \ac{FIM} geometry to both elevation and azimuth angles of every propagation path.
This is a fundamental departure from \ac{RIS}/\ac{SIM} models where electronically controlled phases multiply the response linearly.

\subsection{FIM MIMO Channel Model}

The \ac{MIMO} \ac{DD} channel with \acp{FIM} at both ends, $\mathbf{H}(\bm{y}_\mathrm{T},\bm{y}_\mathrm{R},t,\tau) \in \mathbb{C}^{N_\mathrm{R} \times N_\mathrm{T}}$ can be expressed as
\begin{eqnarray}
\label{eq:MIMO_TD_channel}
\!\!\!\mathbf{H}(\bm{y}_\mathrm{T},\bm{y}_\mathrm{R},t,\tau) \!\triangleq\! \sqrt{\tfrac{N_\mathrm{T}N_\mathrm{R}}{P}} \sum_{p=1}^P h_p e^{\jmath 2\pi \nu_p t} \delta\left(\tau\!-\!\tau_p\right)&&\\
&&\hspace{-32ex}\times  \mathbf{b}_{\mathrm{R}:p}\!\left(\bm{y}_\mathrm{R},\phi_p^{\rm in},\theta_p^{\rm in}\right) \mathbf{b}_{\mathrm{T}:p}\herm\!\left(\bm{y}_\mathrm{T},\phi_p^{\rm out},\theta_p^{\rm out}\right),\nonumber
\end{eqnarray}
where $\mathbf{b}_{\mathrm{T}:p}(\cdot) \in \mathbb{C}^{N_\mathrm{T} \times 1}$ and $\mathbf{b}_{\mathrm{R}:p}(\cdot) \in \mathbb{C}^{N_\mathrm{R} \times 1}$ are the \ac{TX} and \ac{RX} \ac{FIM} \ac{UPA} response vectors from \eqref{eq:steering_vec_FIM}, $h_p$ is the $p$-th complex path gain, $\tau_p \in [0,\tau_{\max}]$ and $\nu_p \in [-\nu_{\max},\nu_{\max}]$ are the delay and Doppler shift, and $(\phi_p^{\rm in},\theta_p^{\rm in})$, $(\phi_p^{\rm out},\theta_p^{\rm out})$ are the \ac{AoA}/\ac{AoD} pairs for the $p$-th path\footnote{While \ac{FIM} motion causes slight perturbations in the path parameters, this effect is negligible when the movement is confined to a factor of $\lambda$ \cite{AnTWC2025}.}.
We define $\tilde{h}_p \triangleq \sqrt{N_\mathrm{T}N_\mathrm{R}/P}\, h_p$ for notational convenience.

\section{I/O Relationships for OFDM, OTFS, and AFDM}
\label{IO_Model}

\subsection{Arbitrarily Modulated Signals}

The system uses $N_\mathrm{T}$ \ac{TX} and $N_\mathrm{R}$ \ac{RX} elements with $d_s \triangleq \min(N_\mathrm{T},N_\mathrm{R})$ data streams.
The baseband received signal at time $t$ is
\begin{equation}
\mathbf{r}(t)
= \int_{-\infty}^\infty \mathbf{H}(\bm{y}_\mathrm{T},\bm{y}_\mathrm{R},t,\tau) \mathbf{s}(t - \tau)\, d\tau + \mathbf{w}(t),
\label{eq:TD_IO}
\end{equation}
where $\mathbf{s}(t) \in \mathbb{C}^{d_s \times 1}$ is the transmit signal and $\mathbf{w}(t)$ is \ac{AWGN} with variance $\sigma_n^2$.

Sampling at rate $F_S = 1/T_S$ and employing a \ac{CP} of length $N_\mathrm{CP}$ with circular convolution, the $v$-th stream discrete-time received vector is \cite{Rou_SPM_2024}
\begin{equation}
\label{eq:vectorized_TD_IO}
\mathbf{r}_v = \sum_{u=1}^{d_s} \underbrace{\sum_{p=1}^P \check{h}_{p,v,u} \overbrace{\mathbf{\Theta}_p \mathbf{\Omega}^{f_p} \mathbf{\Pi}^{\ell_p}}^{\mathbf{G}_p}}_{\bar{\mathbf{H}}_{v,u}(\bm{y}_\mathrm{T},\bm{y}_\mathrm{R})} \mathbf{s}_u + \mathbf{w}_v,
\end{equation}
where $\check{h}_{p,v,u}$ is the $(v,u)$-th element of $\check{\mathbf{H}}_p(\bm{y}_\mathrm{T},\bm{y}_\mathrm{R}) \triangleq \tilde{h}_p \mathbf{b}_{\mathrm{R}:p} \mathbf{b}_{\mathrm{T}:p}\herm$,
$f_p \triangleq N\nu_p/F_s$ and $\ell_p \triangleq \tau_p/T_s$ are normalized Doppler and delay indices.
Here, $\mathbf{\Theta}_p \in \mathbb{C}^{N\times N}$ is a diagonal \ac{CP} phase matrix with entries depending on a waveform-specific phase function $\phi_{\mathrm{CP}}(n)$, $\mathbf{\Omega} \triangleq \text{diag}([1, e^{-\jmath 2\pi/N}, \dots, e^{-\jmath 2\pi(N-1)/N}])$ contains $N$ complex roots of unity, and $\mathbf{\Pi}$ is the $N\times N$ forward cyclic shift matrix.

Using Kronecker products, the overall $Nd_s$-element received signal is
\begin{equation}
\mathbf{r}_\mathrm{TD} = \bar{\mathbf{H}}(\bm{y}_\mathrm{T},\bm{y}_\mathrm{R})\, \mathbf{s}_\mathrm{TD} + \bar{\mathbf{w}}_\mathrm{TD},
\label{eq:vectorized_TD_IO_kron}
\end{equation}
with $\bar{\mathbf{H}}(\bm{y}_\mathrm{T},\bm{y}_\mathrm{R}) \triangleq \sum_{p=1}^P (\check{\mathbf{H}}_p \otimes \mathbf{G}_p) \in \mathbb{C}^{Nd_s \times Nd_s}$.

\subsection{OFDM Signaling}

In \ac{OFDM}, $\mathbf{s}^{(\text{OFDM})}_u = \mathbf{F}_N\herm \mathbf{x}_u$ with $\mathbf{x}_u \in \mathcal{C}^{N\times 1}$.
The \ac{CP} phase function yields $\phi_{\mathrm{CP}}(n)=0$, so $\mathbf{\Theta}_p = \mathbf{I}_N$.
After demodulation via $\mathbf{y}^{(\text{OFDM})}_v = \mathbf{F}_N \mathbf{r}^{(\text{OFDM})}_v$, the effective channel is
\begin{equation}
\label{eq:OFDM_eff}
\bar{\mathbf{H}}_\text{OFDM} = \sum_{p=1}^P \check{\mathbf{H}}_p \otimes \underbrace{(\mathbf{F}_N \mathbf{G}_p \mathbf{F}_N\herm)}_{\mathbf{G}_p^\text{OFDM}}.
\end{equation}

\subsection{OTFS Signaling}

For \ac{OTFS}, symbol matrices $\mathbf{X}_u \in \mathcal{C}^{\tilde{K}\times \tilde{K}'}$ (with $\tilde{K}\tilde{K}'=N$) are modulated via the \ac{IDZT}: $\mathbf{s}^{(\text{OTFS})}_u = (\mathbf{F}_{\tilde{K}'}\herm \otimes \mathbf{I}_{\tilde{K}}) \text{vec}(\mathbf{X}_u)$.
Demodulation uses the \ac{DZT}: $\mathbf{y}^{(\text{OTFS})}_v = (\mathbf{F}_{\tilde{K}'} \otimes \mathbf{I}_{\tilde{K}}) \mathbf{r}^{(\text{OTFS})}_v$.
The \ac{CP} matrices again reduce to identity ($\mathbf{\Theta}_p = \mathbf{I}_N$), and the effective channel is
\begin{equation}
\label{eq:OTFS_eff}
\bar{\mathbf{H}}_\text{OTFS} = \sum_{p=1}^P \check{\mathbf{H}}_p \otimes \underbrace{((\mathbf{F}_{\tilde{K}'} \!\otimes\! \mathbf{I}_{\tilde{K}}) \mathbf{G}_p (\mathbf{F}_{\tilde{K}'}\herm \!\otimes\! \mathbf{I}_{\tilde{K}}))}_{\mathbf{G}_p^\text{OTFS}}.
\end{equation}

\subsection{AFDM Signaling}

\ac{AFDM} modulates via the \ac{IDAFT}: $\mathbf{s}^{(\text{AFDM})}_u = \mathbf{\Lambda}_1\herm \mathbf{F}_{N}\herm \mathbf{\Lambda}_2\herm \mathbf{x}_u$,
where $\mathbf{\Lambda}_i \triangleq \text{diag}([1, e^{-\jmath2\pi c_i 2^2}, \ldots, e^{-\jmath2\pi c_i (N-1)^2}])$ for chirp parameters $c_1, c_2$ \cite{Bemani_TWC_2023}.
The first central chirp frequency $c_1$ is optimized based on maximum Doppler statistics, while $c_2$ can be used for waveform shaping.
Unlike \ac{OFDM}/\ac{OTFS}, the \ac{CP} matrix is replaced by a \ac{CPP} matrix $\bm{\varTheta}_p$ obtained by setting $\phi_\mathrm{CP}(n) = c_1(N^2 - 2Nn)$ in the general \ac{CP} definition.
After \ac{DAFT} demodulation, the effective channel is
\vspace{-1ex}
\begin{equation}
\label{eq:AFDM_eff}
\bar{\mathbf{H}}_\text{AFDM} = \sum_{p=1}^P \check{\mathbf{H}}_p \otimes \underbrace{(\mathbf{\Lambda}_2 \mathbf{F}_{N} \mathbf{\Lambda}_1 \mathbf{G}_p \mathbf{\Lambda}_1\herm \mathbf{F}_{N}\herm \mathbf{\Lambda}_2\herm)}_{\mathbf{G}_p^\text{AFDM}}.
\end{equation}
The shared structure of \eqref{eq:OFDM_eff}--\eqref{eq:AFDM_eff} enables a unified signal processing framework for all three waveforms; see \cite{Rou_SPM_2024} for a detailed comparison.
In what follows, $\bar{\mathbf{G}}_p$ denotes the waveform-dependent matrix for any of the three schemes.

\section{FIM Optimization for ISAC}
\label{sec:FIM_optimization_for_ISAC}

\subsection{Problem Formulation}

We maximize the achievable rate by jointly optimizing the \ac{TX}/\ac{RX} \ac{FIM} shapes $\bm{y}_\mathrm{T},\bm{y}_\mathrm{R}$ and transmit covariance $\mathbf{T} \triangleq \mathbf{x}\mathbf{x}\herm$, subject to morphing range, power, and sensing \ac{QoS} constraints.
The sensing constraint ensures sufficient signal power at target locations for reliable parameter estimation \cite{NiuWCL2024, TVT_2025_Teng_Flexible}.
Using any waveform's effective channel $\bar{\mathbf{H}}(\bm{y}_\mathrm{T},\bm{y}_\mathrm{R})$ from \eqref{eq:OFDM_eff}--\eqref{eq:AFDM_eff}:
\vspace{-1ex}
\begin{subequations}\label{eq:opt_prob}
\begin{alignat}{2}
&\max_{\mathbf{T},\, \bm{y}_\mathrm{T},\, \bm{y}_\mathrm{R}} &\;& \log_{2}\det\!\left( \mathbf{I}_{Nd_s}\!+\!\frac{1}{\sigma_w^{2}}\bar{\mathbf{H}}\mathbf{T}\bar{\mathbf{H}}\herm \right) \label{eq:obj}\\
&\textrm{s.t.} & & \textrm{tr}( \mathbf{T} )\leq P_{\textrm{t}},\; \mathbf{T}\succeq \mathbf{0}, \label{eq:power}\\
& & & \textrm{tr}( \bar{\mathbf{H}}\mathbf{T}\bar{\mathbf{H}}\herm ) \geq \Psi, \label{eq:sensing}\\
& & & y_\text{min} \leq y_{\mathrm{T}, n_\mathrm{T}} \leq y_\text{max},\;\forall n_\mathrm{T}, \label{eq:morph_T}\\
& & & y_\text{min} \leq y_{\mathrm{R}, n_\mathrm{R}} \leq y_\text{max},\;\forall n_\mathrm{R}, \label{eq:morph_R}
\end{alignat}
\end{subequations}
where $\Psi$ is the sensing \ac{QoS} threshold.
The achievable rate in \eqref{eq:obj} serves as a conditional performance metric evaluated with respect to the effective channel: in the \ac{ISAC} setting, channel parameters are inferred through sensing, and the resulting estimate enables adaptive communication.

The sensing constraint \eqref{eq:sensing} is absorbed into the objective via a penalty term $g_c = \min\{ \textrm{tr}(\bar{\mathbf{H}}\mathbf{T}\bar{\mathbf{H}}\herm) - \Psi, 0 \}$ with weight $\beta > 0$.
To focus on the \ac{FIM} geometry impact, we set $\mathbf{T} \approx \mathbf{I}_{Nd_s}$.
This is justified because, under \ac{iid} zero-mean unit-variance transmit symbols across $N$ subcarriers and $d_s$ streams, the normalized covariance $\frac{1}{N}\mathbf{x}\mathbf{x}\herm$ satisfies $\mathbb{E}[\frac{1}{N}\mathbf{x}\mathbf{x}\herm] = \mathbf{I}_{Nd_s}$ and concentrates around $\mathbf{I}_{Nd_s}$ as $Nd_s$ grows large (by the law of large numbers).
While not capacity-optimal in general, this simplification isolates the impact of \ac{FIM} geometry and yields
\vspace{-1ex}
\begin{subequations}\label{eq:reform}
\begin{alignat}{2}
&\max_{\bm{y}_\mathrm{T},\, \bm{y}_\mathrm{R}} &\quad& \log_{2}\det\!\Big( \mathbf{I}_{Nd_s}+\underbrace{\tfrac{1}{\sigma_w^{2}}\bar{\mathbf{H}}\bar{\mathbf{H}}\herm}_{\mathbf{Q}} \Big) + \beta g_c \label{eq:reform_obj}\\
&\textrm{s.t.} & & \eqref{eq:morph_T},\; \eqref{eq:morph_R}.
\end{alignat}
\end{subequations}

\subsection{Proposed Optimization Framework}
\label{subsec:proposed_algorithm}

We solve \eqref{eq:reform} via projected gradient ascent.
Since the objective is non-convex, convergence to a global optimum cannot be guaranteed, but the algorithm empirically converges within ${\sim}10$ iterations.

\subsubsection{Differential of the Effective Channel} Given the vectorized form of the effective channel 
\begin{equation}
    \mathrm{vec} (\bar{\mathbf{H}}) = \mathrm{vec} \Big(  \sum_{p=1}^P \big( \check{\mathbf{H}}_p \otimes {\mathbf{G}}_p \big) \Big),
\end{equation}
it can be rewritten as \cite[Lemma 2.13]{hjorungnes2011}
\begin{equation}
    \label{eq:vec_Hbar}
    \mathrm{vec} (\bar{\mathbf{H}}) = \bm{\Gamma} \sum_{p=1}^P \big( \mathrm{vec} (\check{\mathbf{H}}_p) \otimes \mathrm{vec} ({\mathbf{G}}_p) \big),
\end{equation}
where $\bm{\Gamma} = \mathbf{I}_{N_\mathrm{T}} \otimes \mathbf{C}_{N,N_\mathrm{R}} \otimes \mathbf{I}_{N} $, and $\mathbf{C}_{m,n}$ is the commutation matrix such that $\mathbf{K}_{m,n} \cdot \mathrm{vec}(\mathbf{X}) = \mathrm{vec}(\mathbf{X}\trans)$.
Then, taking the differential of \eqref{eq:vec_Hbar}, which is constant with respect to ${\mathbf{G}}_p$ yields 
\begin{equation}
    \label{eq:dvec_Hbar}
    d\,\mathrm{vec} (\bar{\mathbf{H}}) = \bm{\Gamma} \sum_{p=1}^P \big( d\, \mathrm{vec} (\check{\mathbf{H}}_p) \otimes \mathrm{vec} ({\mathbf{G}}_p) \big).
\end{equation}

\subsubsection{Differential of the $p$-th Contribution}
Given the $p$-th contribution to the equivalent channel $\check{\mathbf{H}}_p(\bm{y}_\mathrm{T},\bm{y}_\mathrm{R}) = \tilde{h}_p \mathbf{b}_{\mathrm{R}:p} \mathbf{b}_{\mathrm{T}:p}\herm$, the identity $\mathrm{vec} (\bm{u}\bm{v}\herm) = ( \bm{v}^* \otimes \mathbf{I} ) \bm{u}$, and \cite[Lemma 3.4]{hjorungnes2011}, its vectorized differential is 
\begin{eqnarray}
    \label{eq:dHp first}
    d \, \mathrm{vec} (\check{\mathbf{H}}_p) = \tilde{h}_p ( \mathbf{b}_{\mathrm{T}:p}^* \otimes \mathbf{I}_{N_\mathrm{R}} ) d\mathbf{b}_{\mathrm{R}:p} \hspace{1ex} \nonumber \\
    + \tilde{h}_p  ( \mathbf{I}_{N_\mathrm{T}} \otimes \mathbf{b}_{\mathrm{R}:p} ) d\mathbf{b}_{\mathrm{T}:p}^* ,
\end{eqnarray}
where the small perturbations of the UPA responses are
\begin{subequations}
\begin{eqnarray}
    \label{eq:db_RT}
    d\,\mathbf{b}_{\mathrm{R}:p} = \frac{\partial \mathbf{b}_{\mathrm{R}:p}}{ \partial \bm{y}_\mathrm{R}} d\, \bm{y}_\mathrm{R} = \gamma_{\mathrm{R}:p} \, \mathrm{diag} (\mathbf{b}_{\mathrm{R}:p}) \, d\bm{y}_\mathrm{R}, \\
    d\,\mathbf{b}_{\mathrm{T}:p}^* = \frac{\partial \mathbf{b}_{\mathrm{T}:p}^*}{ \partial \bm{y}_\mathrm{R}} d\, \bm{y}_\mathrm{T} = \gamma_{\mathrm{T}:p}^* \, \mathrm{diag} (\mathbf{b}_{\mathrm{T}:p}^*) \, d\bm{y}_\mathrm{T},
\end{eqnarray}
\end{subequations}
with the auxiliary variables $ \gamma_{\mathrm{R}:p} \triangleq \jmath \frac{2\pi}{\lambda} \sin\phi_p^{\rm in} \sin\theta_p^{\rm in}  $ and $\gamma_{\mathrm{T}:p} \triangleq \jmath \frac{2\pi}{\lambda} \sin\phi_p^{\rm out} \sin\theta_p^{\rm out}$. Then, plugging \eqref{eq:db_RT} back into \eqref{eq:dHp first} yields
\begin{equation}
    \label{eq:dvec_Hcheck}
    d \, \mathrm{vec} (\check{\mathbf{H}}_p) = \mathbf{K}_{\mathrm{R},p} d\bm{y}_\mathrm{R} + \mathbf{K}_{\mathrm{T},p} d \bm{y}_\mathrm{T},
\end{equation}
where $\mathbf{K}_{\mathrm{R},p} \triangleq \tilde{h}_p \gamma_{\mathrm{R}:p} ( \mathbf{b}_{\mathrm{T}:p}^* \otimes \mathbf{I}_{N_\mathrm{R}} )  \, \mathrm{diag} (\mathbf{b}_{\mathrm{R}:p})$ and $\mathbf{K}_{\mathrm{T},p} \triangleq \tilde{h}_p \gamma_{\mathrm{T}:p}^*  ( \mathbf{I}_{N_\mathrm{T}} \otimes \mathbf{b}_{\mathrm{R}:p} ) \, \mathrm{diag} (\mathbf{b}_{\mathrm{T}:p}^*) $ are auxiliary variables.
Inserting \eqref{eq:dvec_Hcheck} into \eqref{eq:dvec_Hbar} then yields the final differential of the effective channel
\begin{eqnarray}
    d \, \mathrm{vec}(\bar{\mathbf{H}}) = \bm{\Gamma} \sum_{p=1}^P \big( \mathbf{K}_{\mathrm{R},p} \otimes \text{vec} (\mathbf{G}_p) \big)  d \bm{y}_\mathrm{R} \hspace{2ex} \nonumber \\
    + \bm{\Gamma} \sum_{p=1}^P \big( \mathbf{K}_{\mathrm{T},p} \otimes \text{vec} (\mathbf{G}_p) \big) d \bm{y}_\mathrm{T}. 
\end{eqnarray}
%
%
\subsubsection{Tx and Rx Shape Gradients} To maintain tractability, we take the above differentials and convert them to per-element gradients $\bm{y}_{\mathrm{R}:n_t}$ and $\bm{y}_{\mathrm{T}:n_r}$, then separating the contributions from $d \bm{y}_\mathrm{R}$ and $d \bm{y}_\mathrm{T}$, and reversing \cite[Lemma 2.13]{hjorungnes2011} yields
\begin{subequations}
\begin{equation}
    \frac{\partial \bar{\mathbf{H}}}{\partial \bm{y}_{\mathrm{T}:n_t}} = \sum_{p=1}^P \bigg( \tilde{h}_p \gamma_{\mathrm{T}:p}^* \mathbf{b}_{\mathrm{R}:p} (\mathbf{e}_{n_t} \odot \mathbf{b}_{\mathrm{T}:p}) \herm \bigg) \otimes \bar{\mathbf{G}}_p,
    \label{eq:dH_dyT}
\end{equation}
\vspace{-1ex}
\begin{equation}
    \frac{\partial \bar{\mathbf{H}}}{\partial \bm{y}_{\mathrm{R}:n_r}} = \sum_{p=1}^P \bigg( \tilde{h}_p \gamma_{\mathrm{R}:p} (\mathbf{e}_{n_r} \odot \mathbf{b}_{\mathrm{R}:p}) \mathbf{b}_{\mathrm{T}:p}\herm \bigg) \otimes \bar{\mathbf{G}}_p,
    \label{eq:dH_dyR}
\end{equation}
\end{subequations}
with $\mathbf{e}_{n_t} \in \mathbb{C}^{N_\mathrm{T} \times 1}$ denoting the $n_t$-th standard basis vector.

\subsubsection{Objective Gradient} The gradient of $\mathbf{Q}$ w.r.t.\ $\bm{y}_{\mathrm{T}:n_t}$ is
\begin{equation}
    \frac{\partial \mathbf{Q}}{\partial \bm{y}_{\mathrm{T}:n_t}} = \frac{1}{\sigma_w^2} \bigg( \frac{\partial \bar{\mathbf{H}}}{\partial \bm{y}_{\mathrm{T}:n_t}} \bar{\mathbf{H}}\herm + \bar{\mathbf{H}} \Big( \frac{\partial \bar{\mathbf{H}}}{\partial \bm{y}_{\mathrm{T}:n_t}} \Big)\herm \bigg),
    \label{eq:dQ_dyT}
\end{equation}
and the full objective $f \triangleq \log_2\det(\mathbf{I}_{Nd_s}+\mathbf{Q}) + \beta g_c$ has the per-element scalar gradient
\begin{equation}
    \frac{\partial f}{\partial \bm{y}_{\mathrm{T}:n_t}} \!=\! \frac{1}{\ln 2} \Real{\textrm{tr}\!\Big( (\mathbf{I}_{Nd_s} \!+\! \mathbf{Q})^{-1} \frac{\partial \mathbf{Q}}{\partial \bm{y}_{\mathrm{T}:n_t}} \Big)} \!+\! \beta \frac{\partial g_c}{\partial \bm{y}_{\mathrm{T}:n_t}},
    \label{eq:df_dyT}
\end{equation}
where the sensing penalty gradient is
\begin{equation}
    \frac{\partial g_c}{\partial \bm{y}_{\mathrm{T}:n_t}} = \Real{ \textrm{tr}\Big( \frac{\partial \bar{\mathbf{H}}}{\partial \bm{y}_{\mathrm{T}:n_t}} \bar{\mathbf{H}}\herm + \bar{\mathbf{H}} \Big( \frac{\partial \bar{\mathbf{H}}}{\partial \bm{y}_{\mathrm{T}:n_t}} \Big)\herm \Big) }.
    \label{eq:dgc_dyT}
\end{equation}
Identical expressions hold for $\bm{y}_{\mathrm{R}:n_r}$ by replacing TX quantities with their RX counterparts from \eqref{eq:dH_dyR}.

\subsubsection{Surface Shape Update} At each iteration $i$, with step size $\mu > 0$ (typically obtained via Armijo line search):
\begin{subequations}
\label{eq:update}
    \begin{equation}
        \bm{y}_{\mathrm{T}:n_t}^{(i+1)} \leftarrow \mathrm{proj}_{[y_\text{min},y_\text{max}]}\!\Big( \bm{y}_{\mathrm{T}:n_t}^{(i)} + \mu \frac{\partial f}{\partial \bm{y}_{\mathrm{T}:n_t}} \Big),
    \end{equation}
    \vspace{-1.5ex}
    \begin{equation}
        \bm{y}_{\mathrm{R}:n_r}^{(i+1)} \leftarrow \mathrm{proj}_{[y_\text{min},y_\text{max}]}\!\Big( \bm{y}_{\mathrm{R}:n_r}^{(i)} + \mu \frac{\partial f}{\partial \bm{y}_{\mathrm{R}:n_r}} \Big),
    \end{equation}
\end{subequations}
where the projection enforces the constraints \eqref{eq:morph_T}-\eqref{eq:morph_R}.

These shape-gradients \emph{vanish identically} under a purely electronic \ac{RIS}/\ac{SIM} model (where $\bm{y}$ is fixed), confirming that the optimization arises exclusively from \ac{FIM} geometry control.
The complete procedure is summarized in Algorithm~\ref{alg:proposed_decoder}.

\begin{algorithm}[t]
\caption{FIM Optimization for ISAC}
\label{alg:proposed_decoder}
\setlength{\baselineskip}{10.5pt}
\textbf{Input:} Iterations $i_{\mathrm{GD}}$, noise variance $\sigma^2_w$, step size $\mu$, penalty $\beta$.\\
\textbf{Output:} $\bm{y}_\mathrm{T}^\star$, $\bm{y}_\mathrm{R}^\star$.
\begin{algorithmic}[1]
\STATEx \hspace{-3.5ex}\textbf{Initialize:} Random $\bm{y}_\mathrm{T}^\star, \bm{y}_\mathrm{R}^\star$ within $[y_{\min},y_{\max}]$; compute $\bar{\mathbf{H}}^\star$.
\STATEx \hspace{-3.5ex}\textbf{for} $i=1$ to $i_{\mathrm{GD}}$ \textbf{do}:
\STATE Compute $\mathbf{Q}$ from \eqref{eq:reform_obj}.
\STATEx \textbf{for} $n_t=1$ to $N_\mathrm{T}$:
\STATE Compute $\frac{\partial \bar{\mathbf{H}}}{\partial \bm{y}_{\mathrm{T}:n_t}}$ via \eqref{eq:dH_dyT}, then $\frac{\partial \mathbf{Q}}{\partial \bm{y}_{\mathrm{T}:n_t}}$ via \eqref{eq:dQ_dyT}.
\STATE Compute $\frac{\partial f}{\partial \bm{y}_{\mathrm{T}:n_t}}$ via \eqref{eq:df_dyT}-\eqref{eq:dgc_dyT}. 
\STATEx \textbf{end for}
\STATEx \textbf{for} $n_r=1$ to $N_\mathrm{R}$:
\STATE Compute RX gradients analogously using \eqref{eq:dH_dyR}. \STATEx \textbf{end for}
\STATE Update surface shapes via \eqref{eq:update}.
\STATEx \hspace{-3.5ex}\textbf{end for}
\end{algorithmic}
\end{algorithm}

\section{Performance Analysis}
\label{sec:performance_analysis}

Unless otherwise specified, parameters in Table~\ref{tab:simulation_parameters} are used throughout.
For the penalty factor, $\beta = 2$ provides a reasonable communication-sensing tradeoff \cite{NiuWCL2024}.

\subsection{Communications Performance}

\begin{figure}[h]
\subfigure[{\footnotesize $N = 16$, $P = 2$}]%
{\includegraphics[width=\columnwidth]{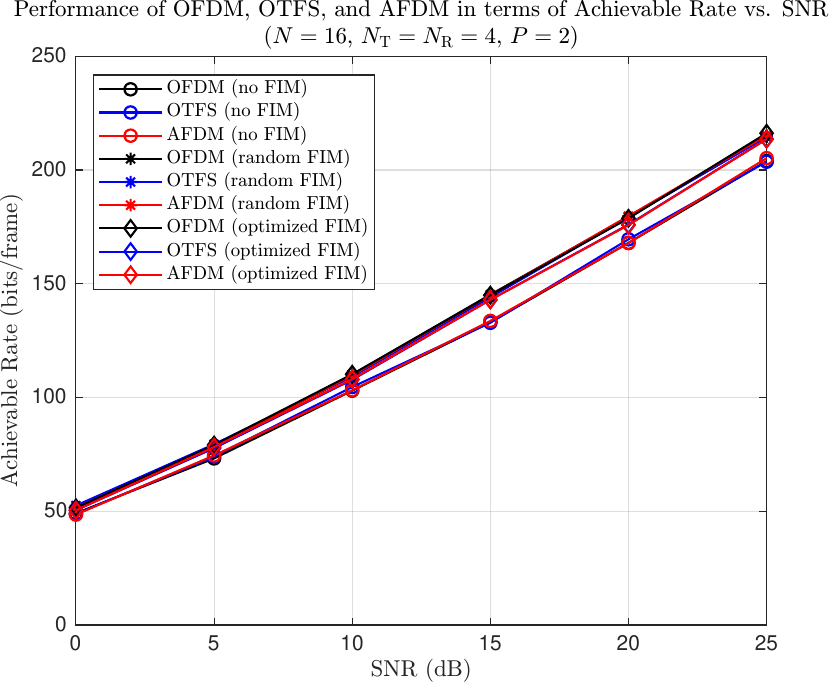}
\label{fig:comm_N16}}\\[-1ex]
\subfigure[{\footnotesize $N = 64$, $P = 5$}]%
{\includegraphics[width=\columnwidth]{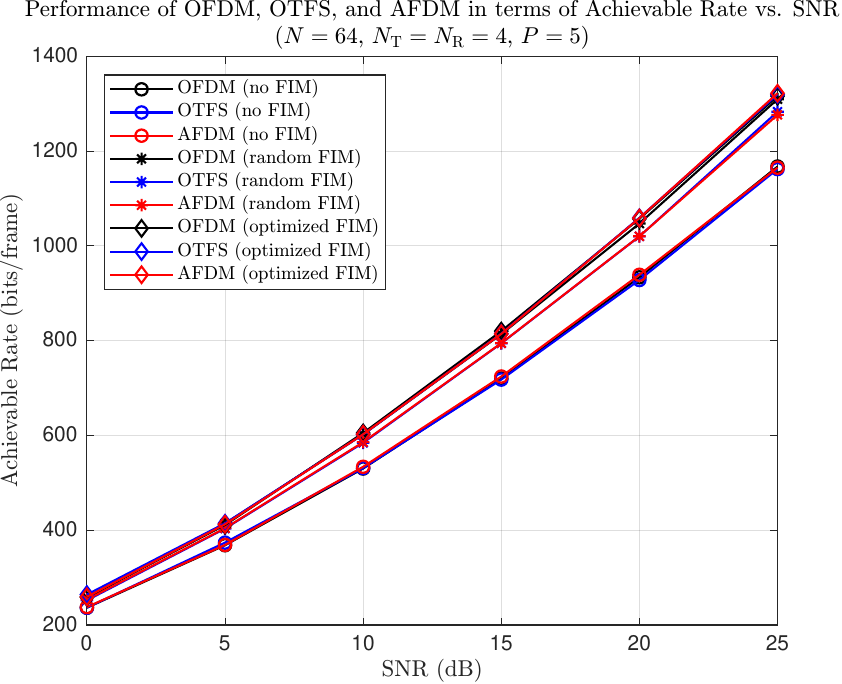}
\label{fig:comm_N64}}
\vspace{-2ex}
\caption{Achievable Rate vs.\ \ac{SNR} for the proposed \ac{FIM} optimization.}
\label{fig:rate_plots_vs_SNR}
\vspace{-2ex}
\end{figure}

Figure~\ref{fig:comm_N16} presents the achievable rate versus \ac{SNR} for \ac{OFDM}, \ac{OTFS}, and \ac{AFDM} under three cases: no \acp{FIM}, randomly tuned \acp{FIM}, and \acp{FIM} optimized via Algorithm~\ref{alg:proposed_decoder}.
The key insight is that \acp{FIM} significantly improve the achievable rate: a gain of ${\sim}2.5$\,dB from no \acp{FIM} to random \acp{FIM}, and an additional ${\sim}2$\,dB from random to optimized \acp{FIM}.
The achievable rates across waveforms are nearly identical, since all operate within the same physical channel's bandwidth and \ac{SNR} constraints---this is consistent with Shannon's capacity formula.

\begin{figure}[h]
\subfigure[{\footnotesize Elevation Angle Profile}]%
{\includegraphics[width=\columnwidth]{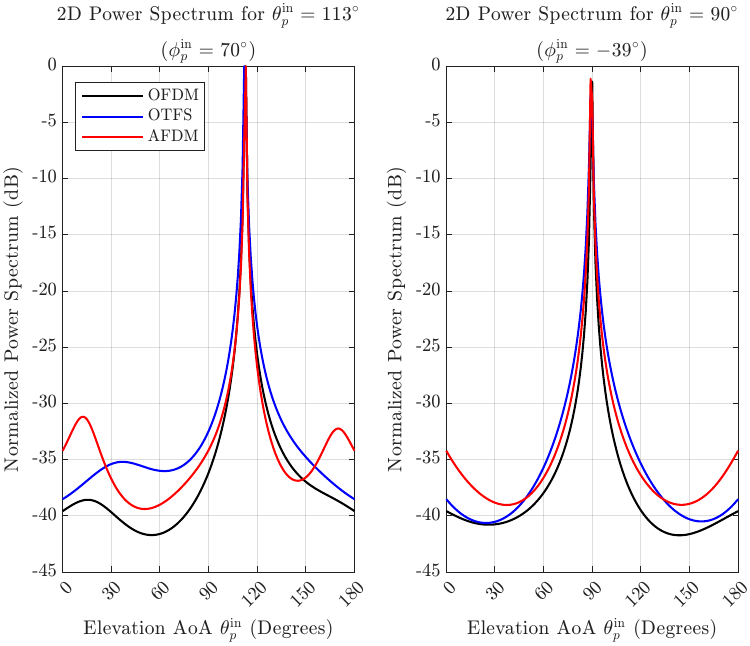}
\label{fig:Sens_2D_e}}\\[-1ex]
\subfigure[{\footnotesize Azimuth Angle Profile}]%
{\includegraphics[width=\columnwidth]{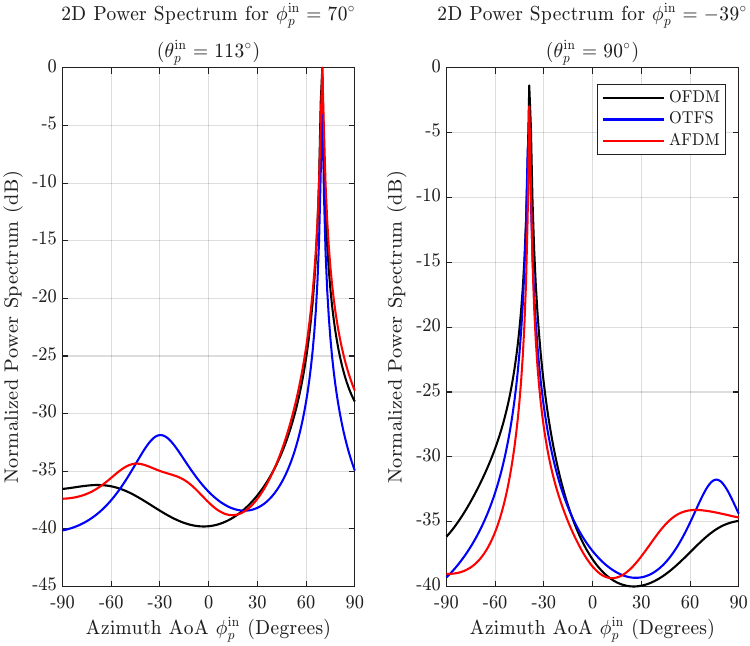}
\label{fig:Sens_2D_a}}
\vspace{-2ex}
\caption{Normalized \ac{2D} \ac{MUSIC} power spectrum for $P\!=\!2$ scatterers.}
\label{fig:Sens_2D}
\vspace{-2ex}
\end{figure}

\begin{table}[h]
\centering
\caption{System Parameters}
\vspace{-1ex}
\label{tab:simulation_parameters}
\begin{tabular}{|c|c|c|}
\hline
\textbf{Parameter} & \textbf{Symbol} & \textbf{Value} \\
\hline
Carrier Frequency & $f_c$ & 28 GHz \\
\hline
System Bandwidth & $F_S$ & 20 MHz \\
\hline
Subcarriers & $N$ & 16, 64 \\
\hline
TX/RX-FIM elements & $N_\mathrm{T}$, $N_\mathrm{R}$ & 4 (2$\times$2) \\
\hline
RF chains & $d_s$ & 4 \\
\hline
Channel Scatterers & $P$ & 2, 5 \\
\hline
Morphing Range & $[y_\text{min}, y_\text{max}]$ & $[-\lambda, \lambda]$ \\
\hline
Penalty Factor & $\beta$ & 2 \\
\hline
\end{tabular}
\vspace{-2ex}
\end{table}

However, practical \ac{BER} performance differs: \ac{AFDM} and \ac{OTFS} consistently outperform \ac{OFDM} in high-mobility scenarios due to their inherent \ac{ICI} resilience \cite{RanasingheARXIV2024}.
For the larger system ($N\!=\!64$, $P\!=\!5$, Figure~\ref{fig:comm_N64}), the same trends hold with amplified gains due to increased \acp{DoF}.

\subsection{Sensing Performance}

For sensing, we consider a bistatic setup where the \ac{RX}-\ac{FIM} estimates the elevation and azimuth \acp{AoA} of all $P$ scatterers using the \ac{2D} \ac{MUSIC} algorithm \cite{Ranasinghe_ICASSP_2024}.
The received signal is preprocessed as $\bar{\mathbf{Y}} = \text{vec}^{-1}(\bar{\mathbf{y}})\herm \in \mathbb{C}^{d_s \times N}$, from which the covariance $\mathbf{R}_{\bar{\mathbf{y}}} = \bar{\mathbf{Y}} \bar{\mathbf{Y}}\herm$ is computed.
As shown in Appendix~\ref{app:covariance_derivation}, under standard assumptions this covariance decomposes into $\mathbf{R}_{\bar{\mathbf{y}}} \approx \sum_{p=1}^P \sigma_{s,p}^2\,\mathbf{b}_{\mathrm{R}:p}\mathbf{b}_{\mathrm{R}:p}\herm + \sigma_w^2\mathbf{I}_{d_s}$, enabling standard subspace methods.
The noise subspace eigenvectors $\bar{\mathbf{U}}_N$ yield the \ac{MUSIC} spectrum:
\begin{equation}
    M(\phi,\theta) = \frac{1}{\mathbf{b}_{\mathrm{R}}\herm(\bm{y}_\mathrm{R}^\star,\phi,\theta)\, \bar{\mathbf{U}}_N \bar{\mathbf{U}}_N\herm\, \mathbf{b}_{\mathrm{R}}(\bm{y}_\mathrm{R}^\star,\phi,\theta)}.
\end{equation}

Figure~\ref{fig:Sens_2D} shows the \ac{2D} \ac{MUSIC} spectrum cuts for $P\!=\!2$ scatterers.
With optimized \acp{FIM}, the spectrum clearly identifies both scatterers' \acp{AoA} with strong, well-isolated peaks.
With randomly tuned \acp{FIM}, the peaks are present but fail to correctly isolate scatterers, and without \acp{FIM}, multiple spurious peaks appear.
All three waveforms show similar peak power, with \ac{OFDM} exhibiting slightly lower sidelobes.

\section{Conclusion}
\label{sec:conclusion}

We proposed a \ac{FIM}-based \ac{ISAC} \ac{DD} \ac{MIMO} channel model for high-mobility scenarios and derived unified \ac{I/O} relationships for \ac{OFDM}, \ac{OTFS}, and \ac{AFDM}.
A gradient ascent algorithm with closed-form shape-gradients was developed to maximize achievable rate under sensing constraints.
Numerical results confirm that \ac{FIM} geometry morphing yields significant gains in both communications and sensing performance, underscoring the importance of precise metasurface parametrization for robust \ac{ISAC} across all evaluated waveforms.
Future work includes extension to multi-user scenarios, \ac{FIM}-enhanced angular resolution analysis, and ambiguity function characterization with \ac{FIM} parametrization.

\appendices
\section{Derivation of the 2D MUSIC Covariance Matrix}
\label{app:covariance_derivation}

Starting from the received signal $\bar{\mathbf{y}} = \sum_{p=1}^{P} (\tilde{h}_p\,\mathbf{b}_{\mathrm{R}:p}\mathbf{b}_{\mathrm{T}:p}\herm \otimes \mathbf{G}_p)\,\mathbf{x} + \bar{\mathbf{w}}$, we extract the $n$-th snapshot $\bar{\mathbf{y}}[n] = (\mathbf{I}_{d_s}\otimes\mathbf{e}_n\trans)\bar{\mathbf{y}}$.
Via the Kronecker mixed-product property, this simplifies to $\bar{\mathbf{y}}[n] = \sum_{p=1}^P \mathbf{b}_{\mathrm{R}:p}\, s_p[n] + \bar{\mathbf{w}}[n]$, where $s_p[n] \triangleq \tilde{h}_p\,\mathbf{e}_n\trans\mathbf{G}_p(\mathbf{b}_{\mathrm{T}:p}\herm\otimes\mathbf{I}_N)\,\mathbf{x}$ is a scalar signal term.

The exact covariance $\mathbf{R}_{\bar{\mathbf{y}}} = \mathbb{E}[\bar{\mathbf{y}}[n]\bar{\mathbf{y}}[n]\herm]$ can be shown (under $\mathbb{E}[\mathbf{x}\mathbf{x}\herm]=\sigma_x^2\mathbf{I}_{Nd_s}$, $\mathbf{x}\perp\!\!\perp\bar{\mathbf{w}}$) to equal
$\sigma_x^2 \sum_{p,q} \tilde{h}_p\tilde{h}_q^{*}
(\mathbf{b}_{\mathrm{T}:p}\herm\mathbf{b}_{\mathrm{T}:q})
(\mathbf{e}_n\trans\mathbf{G}_p\mathbf{G}_q\herm\mathbf{e}_n^{*})
\mathbf{b}_{\mathrm{R}:p}\mathbf{b}_{\mathrm{R}:q}\herm
+ \sigma_w^2\mathbf{I}_{d_s}$.
The cross-terms ($p \neq q$) vanish when either (i) $\mathbf{b}_{\mathrm{T}:p}\herm\mathbf{b}_{\mathrm{T}:q} = 0$ (orthogonal transmit steering) or (ii) $\mathbf{e}_n\trans\mathbf{G}_p\mathbf{G}_q\herm\mathbf{e}_n^{*} = 0$ (per-snapshot waveform orthogonality), conditions met in practice via sufficient angular separation or orthogonal subcarriers/chirps.
This yields $\mathbf{R}_{\bar{\mathbf{y}}} \approx \sum_{p=1}^P \sigma_{s,p}^2\,\mathbf{b}_{\mathrm{R}:p}\mathbf{b}_{\mathrm{R}:p}\herm + \sigma_w^2\mathbf{I}_{d_s}$, the standard signal-plus-noise form required by \ac{MUSIC}.


\begin{thebibliography}{00}

\bibitem{RanasingheTWCFIM}
K.~R.~R. Ranasinghe \emph{et al.}, ``Flexible Intelligent Metasurfaces in High-Mobility MIMO Integrated Sensing and Communications,'' \emph{IEEE Trans. Wireless Commun.}, vol.~25, 2026.

\bibitem{GiordaniCOMMAG2020}
M.~Giordani \emph{et al.}, ``Toward 6G networks: Use cases and technologies,'' \emph{IEEE Commun. Mag.}, vol.~58, no.~3, pp. 55--61, 2020.

\bibitem{Bliss_Govindasamy_2013}
D.~W. Bliss and S.~Govindasamy, \emph{Adaptive Wireless Communications}.\hskip 1em Cambridge Univ. Press, 2013.

\bibitem{liu2022integrated}
F.~Liu \emph{et al.}, ``Integrated sensing and communications: Toward dual-functional wireless networks for 6G and beyond,'' \emph{IEEE J. Sel. Areas Commun.}, vol.~40, no.~6, pp. 1728--1767, 2022.

\bibitem{Rou_SPM_2024}
H.~S. Rou \emph{et al.}, ``From OTFS to AFDM: A comparative study of next-generation waveforms for ISAC in doubly dispersive channels,'' \emph{IEEE Signal Process. Mag.}, vol.~41, no.~5, 2024.

\bibitem{RanasingheARXIV2024}
K.~R.~R. Ranasinghe \emph{et al.}, ``Joint channel, data, and radar parameter estimation for AFDM systems in doubly-dispersive channels,'' \emph{IEEE Trans. Wireless Commun.}, vol.~24, no.~2, 2025.

\bibitem{AlexandropolousVTM2024}
G.~C. Alexandropoulos \emph{et al.}, ``Hybrid reconfigurable intelligent metasurfaces for 6G,'' \emph{IEEE Veh. Technol. Mag.}, vol.~19, no.~1, pp. 75--84, 2024.

\bibitem{AnWC2024}
J.~An \emph{et al.}, ``Stacked intelligent metasurface-aided MIMO transceiver design,'' \emph{IEEE Wireless Commun.}, vol.~31, no.~4, pp. 123--131, 2024.

\bibitem{TWC_2024_An_Flexible}
J.~An \emph{et al.}, ``Flexible intelligent metasurfaces for downlink multiuser MISO communications,'' \emph{IEEE Trans. Wireless Commun.}, vol.~24, no.~4, pp. 2940--2955, 2025.

\bibitem{Nature_2022_Bai_A}
Y.~Bai \emph{et al.}, ``A dynamically reprogrammable surface with self-evolving shape morphing,'' \emph{Nature}, vol.~609, pp. 701--708, 2022.

\bibitem{TCOM_2025_An_Flexible}
J.~An \emph{et al.}, ``Flexible intelligent metasurfaces for enhancing MIMO communications,'' \emph{IEEE Trans. Commun.}, vol.~73, no.~9, pp. 7349--7365, 2025.

\bibitem{TVT_2025_Teng_Flexible}
Z.~Teng \emph{et al.}, ``Flexible intelligent metasurface for enhancing multi-target wireless sensing,'' \emph{IEEE Trans. Veh. Technol.}, pp. 1--6, 2025.

\bibitem{ranasinghe2025metasurfacesintegrateddoublydispersivemimochannel}
K.~R.~R. Ranasinghe \emph{et al.}, ``Metasurfaces-integrated doubly-dispersive MIMO: Channel modeling and optimization,'' 2025. [Online]. Available: \texttt{arxiv.org/abs/2506.14985}

\bibitem{ranasinghe2025doublydispersivemimochannelsstacked}
K.~R.~R. Ranasinghe \emph{et al.}, ``Doubly-Dispersive MIMO Channels With Stacked Intelligent Metasurfaces: Modeling, Parametrization, and Receiver Design,'' \emph{IEEE Trans. Wireless Commun.}, 2025.

\bibitem{AnTWC2025}
J.~An, C.~Yuen \emph{et al.}, ``Flexible intelligent metasurfaces for downlink multiuser MISO communications,'' \emph{IEEE Trans. Wireless Commun.}, 2025.

\bibitem{Bemani_TWC_2023}
A.~Bemani, N.~Ksairi, and M.~Kountouris, ``Affine frequency division multiplexing for next generation wireless communications,'' \emph{IEEE Trans. Wireless Commun.}, vol.~22, no.~11, pp. 8214--8229, 2023.

\bibitem{NiuWCL2024}
H.~Niu \emph{et al.}, ``Stacked intelligent metasurfaces for integrated sensing and communications,'' \emph{IEEE Wireless Commun. Lett.}, 2024.

\bibitem{hjorungnes2011}
Hjørungnes A. ``Complex-Valued Matrix Derivatives: With Applications in Signal Processing and Communications,'' \emph{Cambridge Univ. Press}, 2011.

\bibitem{Ranasinghe_ICASSP_2024}
K.~R.~R. Ranasinghe, H.~S. Rou, and G.~T.~F. de~Abreu, ``Fast and efficient sequential radar parameter estimation in MIMO-OTFS systems,'' in \emph{Proc. IEEE ICASSP}, 2024.

\bibitem{Hadani_WCNC_2017}
R.~Hadani \emph{et al.}, ``Orthogonal time frequency space modulation,'' in \emph{Proc. IEEE WCNC}, 2017.

\bibitem{cheng2022integrated}
X.~Cheng \emph{et al.}, ``Integrated sensing and communications for vehicular communication networks,'' \emph{IEEE Internet Things J.}, vol.~9, no.~23, pp. 23\,441--23\,451, 2022.

\end{thebibliography}
\end{document}